SUPERCONDUCTORS

# Evidence for orbital order and its relation to superconductivity in FeSe$_{0.4}$Te$_{0.6}$

Udai R. Singh,[1] Seth C. White,[1] Stefan Schmaus,[1] Vladimir Tsurkan,[2,3] Alois Loidl,[2] Joachim Deisenhofer,[2] Peter Wahl[1,4]*

The emergence of nematic electronic states accompanied by a structural phase transition is a recurring theme in many correlated electron materials, including the high-temperature copper oxide– and iron-based superconductors. We provide evidence for nematic electronic states in the iron-chalcogenide superconductor FeSe$_{0.4}$Te$_{0.6}$ from quasi-particle scattering detected in spectroscopic maps. The symmetry-breaking states persist above $T_c$ into the normal state. We interpret the scattering patterns by comparison with quasi-particle interference patterns obtained from a tight-binding model, accounting for orbital ordering. The relation to superconductivity and the influence on the coherence length are discussed.

## INTRODUCTION

The crystal structure of iron-based superconductors, similar to that of cuprate-based high-temperature superconductors, consists of quasi–two-dimensional iron-pnictide or iron-chalcogenide layers with an iron square lattice at their center (1, 2). Some of the materials undergo a structural phase transition to an orthorhombic phase in the iron pnictides accompanied by a magnetic phase transition, breaking $C_4$ symmetry (3). In both iron-based superconductors and cuprates, the symmetry-breaking states emerge on the underdoped side of the phase diagram, where superconductivity is suppressed and the transition to a magnetically ordered phase takes place (3–6). Nematic electronic states have also been detected in systems that exhibit a quantum critical point, indicating an intimate relationship between electronic correlations and nematic ordering (7, 8). With a number of techniques [for example, scanning tunneling microscopy (STM) (4, 9, 10), angle-resolved photoemission spectroscopy (ARPES) (11), optical spectroscopy (12), and transport (13)], anisotropy in the electronic states, which appears far too large to be explained in full by the orthorhombic distortion, has been found. In BaFe$_2$(As$_{1-x}$P$_x$)$_2$, torque magnetometry shows signatures of electronic anisotropy at temperatures above the structural and magnetic phase transition (14). Specifically, the surprisingly strong resistivity anisotropy observed in Co-doped iron-based superconductors (15) has been successfully described by anisotropic scattering at impurities (16), which locally pin nematic fluctuations (17). The role of electronic nematicity in the context of superconductivity and its relation to the magnetostructural phase transition found in iron-based superconductors remain unclear. Theoretically, the anisotropies in transport and the electronic structure can be accounted for by including symmetry breaking via orbital or magnetic order (18–20). Orbital order leads to a reconstruction of the Fermi surface, as does magnetic order (21).

The key question is: what is the role of these nematic states with respect to superconductivity? Although there is growing evidence that magnetic fluctuations are responsible for the emergence of superconductivity in the iron-based materials (22), it can also be mediated through a coupling of Cooper pairs by orbital fluctuations, albeit with a different order parameter (23). Even if they are not the dominant coupling mechanism, orbital fluctuations have been shown to strengthen pairing in spin fluctuation–mediated coupling (24). Both can lead to anisotropy once static order sets in.

The iron-chalcogenide superconductor FeSe$_{0.4}$Te$_{0.6}$ becomes superconducting at $T_c \approx 14$ K (25) and maintains tetragonal symmetry in both the normal and the superconducting state down to 4 K; no phase transition from a tetragonal to an orthorhombic lattice structure has been reported (26, 27), neither has static magnetic order been found (27). In neutron scattering, low-energy spin excitations are observed at a wave vector (1/2, 1/2, $L$) with a resonance consistent with $s^{\pm}$ pairing (28, 29), which is the same as that in iron-pnictide superconductors (30) and different from the magnetic ordering vector in Fe$_{1+y}$Te (27, 31).

Here, we present a quasi-particle interference (QPI) study of a single crystal of FeSe$_{0.4}$Te$_{0.6}$ by STM. We illustrate breaking of $C_4$ symmetry in both the superconducting and normal states, and discuss its role in the formation of Cooper pairs at low temperatures. Further, we compare our results to QPI patterns calculated from a tight-binding model. We discuss possible origins of the symmetry breaking.

## RESULTS

Among the iron-based superconductors, the iron chalcogenides are particularly well suited for STM experiments: they have a well-defined cleavage plane and expose a nonpolar surface, and from low-energy electron diffraction (32, 33) and ARPES (33), there is no indication of a surface reconstruction or the formation of surface states. STM topographies of our sample (for details of growth and characterization, see Materials and Methods and fig. S1) recorded after cleaving at low temperature show a surface as in Fig. 1A, yielding atomic resolution. The surface layer imaged by STM consists of the chalcogen ions selenium and tellurium that are imaged with different apparent heights, in agreement with previous STM studies (33, 34). The ionic radii of selenium and tellurium suggest that atoms imaged higher (brighter) are tellurium atoms and atoms imaged lower (darker) are selenium atoms (35). This assignment is consistent with a composition analysis (Fig. 1D, fig. S2, and note S1) on the basis of the apparent heights, which yields a composition within a few atomic percent of the one determined from energy-dispersive x-ray (EDX) measurements (36). We do not observe excess Fe atoms in STM images of this sample, which normally show up as protrusions in the topography and have distinct spectroscopic signatures.

[1]Max-Planck-Institut für Festkörperforschung, Heisenbergstrasse 1, D-70569 Stuttgart, Germany. [2]Center for Electronic Correlations and Magnetism, Experimental Physics V, University of Augsburg, D-86159 Augsburg, Germany. [3]Institute of Applied Physics, Academy of Sciences of Moldova, MD 2028 Chisinau, Republic of Moldova. [4]Scottish Universities Physics Alliance, School of Physics and Astronomy, University of St. Andrews, North Haugh, St. Andrews, Fife KY16 9SS, UK.
*Corresponding author. E-mail: wahl@st-andrews.ac.uk

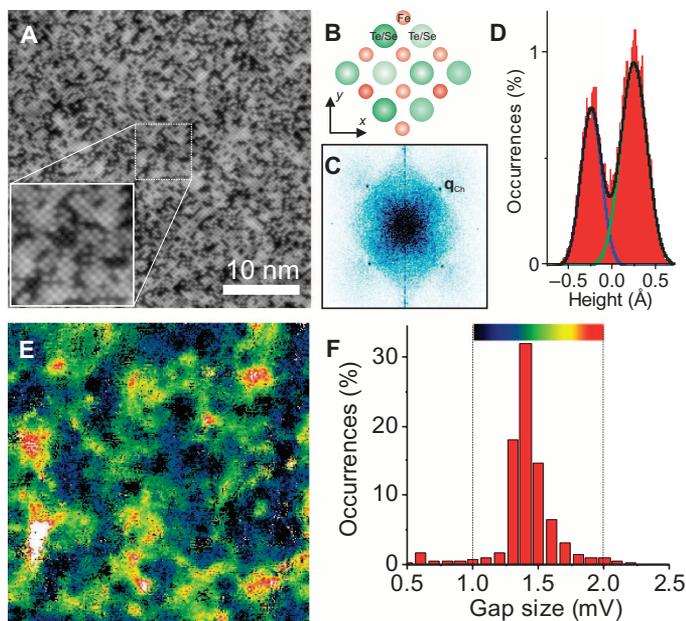

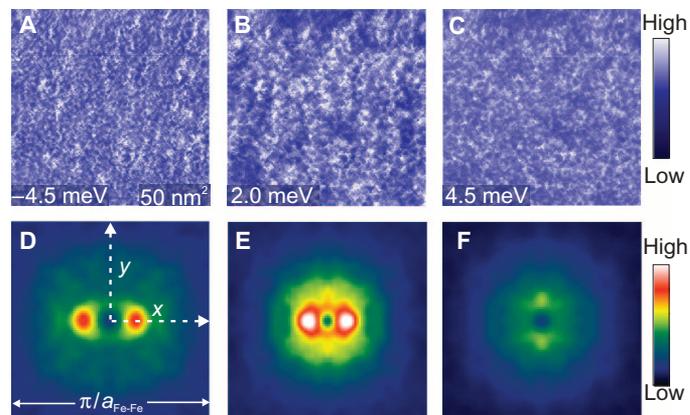

**Fig. 1. Composition analysis from the topography and spatial map of the superconducting gap.** (**A**) STM topography of FeSe$_{0.4}$Te$_{0.6}$ taken with bias voltage $V = 80$ mV and tunneling current $I = 2.0$ nA at 2.1 K. The inset is a close-up of the atomically resolved chalcogenide layer. (**B**) Schematic of the atomic structure of Fe(Se,Te), oriented as in the topography (not to scale). (**C**) Fourier transform of the topographic image shown in (A). Bragg peaks associated with tellurium/selenium ($q_{Ch}$) atoms are visible. (**D**) Histogram of the height of Se/Te atoms in (A); the composition is obtained from two Gaussians fitted to the histogram. The result (37 ± 4% Se and 63 ± 4% Te) closely matches the composition determined by EDX (see also fig. S2 and note S1). (**E**) Spatial map of the size of the superconducting gap in an area of 50 × 50 nm$^2$, taken at $T = 2.1$ K. (**F**) Histogram of the gap size for the map in (E).

**Fig. 2. Symmetry-breaking QPI in FeSe$_{0.4}$Te$_{0.6}$.** (**A** to **C**) Spectroscopic maps of the differential conductance $g(\mathbf{x}, V)$ taken at $T = 2.1$ K (50 × 50 nm$^2$, $V = 40$ mV, $I = 0.3$ nA, $V_{mod} = 600$ μV). (**D** to **F**) Processed Fourier transform images $\tilde{g}(\mathbf{q}, V)$ obtained from (A) to (C) (see note S2 for details of data processing).

Fourier transformation of the topographic image (Fig. 1C) shows no sign of $C_4$ symmetry breaking. Spatial maps of the superconducting gap (Fig. 1E) show a gap inhomogeneity on the order of 20% (Fig. 1F) (*36*).

Spectroscopic maps acquired at a temperature of $T = 2.1$ K over an extended surface area allow us to investigate the electronic structure in both real and momentum space. The dominant scattering vectors, which reveal information about the electronic structure of the material, can be extracted by analyzing the Fourier components of constant energy cuts through these measurements. Spatial maps of the differential conductance $g(\mathbf{x}, V)$ (Fig. 2, A to C) can be taken as proportional to the local density of states. In Fig. 2 (D to F), the Fourier transforms $\tilde{g}(\mathbf{q}, V)$ of the data in Fig. 2 (A to C) are shown. We observe a clear difference in the scattering patterns between the two nearest-neighbor Fe-Fe directions (that is along the *x* and *y* axes in Fig. 2D), which are equivalent for a tetragonal crystal structure. The symmetry breaking becomes more pronounced at negative bias voltages. Its characteristic wave vector shows only weak dispersion across the Fermi energy. This can be seen in more detail from line cuts taken along the nearest-neighbor Fe-Fe directions (Fig. 3, A and B), in which we can trace the symmetry-breaking states (bright peaks) as a function of energy. The wave vector indicates that the unidirectional modulation along the *x* axis occurs with an approximate periodicity of 4 nm (see also fig. S3, where the same conclusion is reached from an analysis of the autocorrelation).

To clarify the relation of these symmetry-breaking states to superconductivity, we have taken maps at a temperature of $T = 16$ K ($> T_c$). The data show that the symmetry-breaking excitations persist into the normal state, and superconductivity forms on top of these excitations. In a line cut through the Fourier transform of differential conductance maps, it can be seen that the characteristic wave vector of the symmetry-breaking states, which becomes gapped in the low-temperature measurements because of superconductivity, can now be traced across the Fermi energy (Fig. 3, C and D, and fig. S4). We show a quantitative analysis of the dominant scattering vector in Fig. 3E. The magnitude of this vector is

$$|\mathbf{q}| \approx \frac{0.12\pi}{a_{\text{Fe-Fe}}} = 2\pi/16 a_{\text{Fe-Fe}}$$

along the nearest-neighbor Fe-Fe direction below and above $T_c$; it slightly decreases with increasing energy. A similar anisotropic scattering vector has been observed in the orthorhombic phase in thin films of FeSe by STM (*37*) and recently in bulk FeSe (*38*). The magnitude of the dominant symmetry-breaking wave vector is substantially smaller than what has been observed in 122 compounds (*4*).

An analysis of the autocorrelation of spatial maps of the superconducting gap as in Fig. 1E reveals strong anisotropy in the decay length of the correlation coefficient in a direction parallel or normal to the wave vector of the symmetry-breaking excitations (Fig. 4A). The decay length of the autocorrelation is a measure for the superconducting coherence length. It can be seen that the main directions coincide with those of the symmetry-breaking modulations. Fits of an exponential decay to horizontal and vertical line cuts yield characteristic length scales of $\xi_x = 1.38$ nm and $\xi_y = 3.0$ nm (Fig. 4B). The superconducting coherence length is suppressed in the direction perpendicular to the stripe-like modulations (parallel to their wave vector), whereas it is enhanced parallel to them. The interplay between nematic order and superconductivity, as well as the anisotropic behavior of the superconducting coherence length, has been studied theoretically, showing that a small anisotropy in the hopping parameters can lead to substantial anisotropy in the coherence length (*39*, *40*). The anisotropy of the coherence length that we observe is similar to the one found near vortex cores in FeSe (*37*), with the difference that FeSe has an orthorhombic crystal structure, which breaks $C_4$ symmetry.

To investigate the relation between the symmetry-breaking states and spatial inhomogeneity of the superconducting gap, we show in Fig. 4C the correlation between a low-pass filtered map obtained at a bias voltage where the symmetry-breaking excitations are visible (see also fig. S5A and note S3) and a gap map obtained in the same field of view (Fig. 1E), revealing an anticorrelation between the two: where the intensity in the low-pass filtered map is high, the local superconducting gap size tends to be smaller and vice versa. Comparison of a gap map to a map of the strength of the order parameter of the symmetry-breaking states reveals a far smaller and less reproducible correlation (fig. S5, B and C).

We can model the anisotropy in the QPI pattern by considering a tight-binding model (note S4). We use the five-band model by Graser et al. (41), with appropriate renormalizations of the bands to ensure consistency with ARPES measurements (33). We include orbital splitting by imposing different occupations for the $d_{xz}$ and $d_{yz}$ orbitals, which leads to a ferro-orbital ordering (21, 42, 43). A tight-binding model including orbital ordering has been previously used to describe the anisotropy of vortex cores in FeSe (40). We have not considered magnetic ordering because there is no evidence for static magnetic order in $FeSe_xTe_{1-x}$ through a wide range of the phase diagram (26, 44), apart from pure FeTe (27, 31). Figure 5 (A and B) shows the Fermi surface calculated from the tight-binding model with tetragonal symmetry and the associated joint density of states (JDOS) to model the QPI pattern. Lifting the degeneracy between the $d_{xz}$ and $d_{yz}$ orbitals by imposing an orbital splitting leads to a distortion of the Fermi surface (Fig. 5C) and consequently to anisotropic QPI scattering (Fig. 5D). The dominant scattering vectors with orbital splitting are the two pronounced maxima near $\mathbf{q} = 0$. This conclusion remains valid when accounting for the orbital character in the QPI calculation (see note S5 and fig. S7). The magnitude of this prominent scattering vector sensitively depends on the amount of orbital splitting introduced, although it shows only a little dispersion (compare fig. S6C). It can therefore be used to estimate the amount of orbital splitting (see Fig. 5E). Comparison to the $\mathbf{q}$ vector of the symmetry-breaking states in our data, which is $|\mathbf{q}| \approx 0.12\pi/a_{Fe-Fe}$, allows us to estimate the amount of orbital splitting to 8 meV (marked by a green arrow in Fig. 5E), substantially less than the 60 meV that has been extracted for the 122 compounds (11). This orbital splitting would correspond to a temperature scale of 50 K. The precise value sensitively depends on the band renormalization. In a recent ARPES study, the opening of a gap near the Γ point has been reported, which can be interpreted in terms of an orbital splitting of ~18 meV for $FeTe_{0.55}Se_{0.45}$ (45), somewhat larger than what we found.

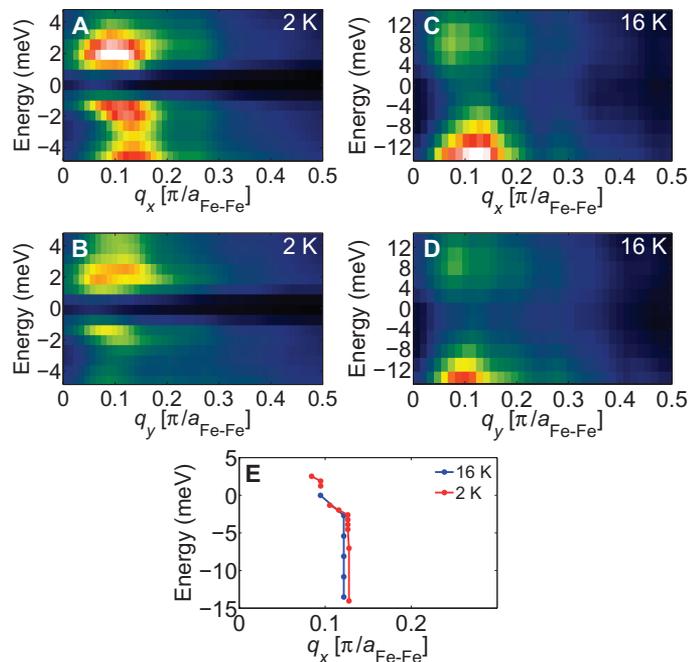

**Fig. 3. Analysis of QPI.** (**A** and **B**) Line cuts in horizontal and vertical directions from the center in Fig. 2 (D to F). A clear anisotropy is seen at negative bias voltages, where the horizontal cut along the Fe-Fe direction shows strong scattering at $|\mathbf{q}| = 0.12\pi/a_{Fe-Fe}$, where $a_{Fe-Fe}$ is the atomic distance between two Fe atoms. (**C** and **D**) Line cuts as in (A) and (B) obtained from a map measured at $T = 16$ K $> T_c$ (see also fig. S4). (**E**) Magnitude $|\mathbf{q}|$ of the dominant scattering vector due to the symmetry-breaking state as a function of energy. The symmetry-breaking excitations persist above $T_c$.

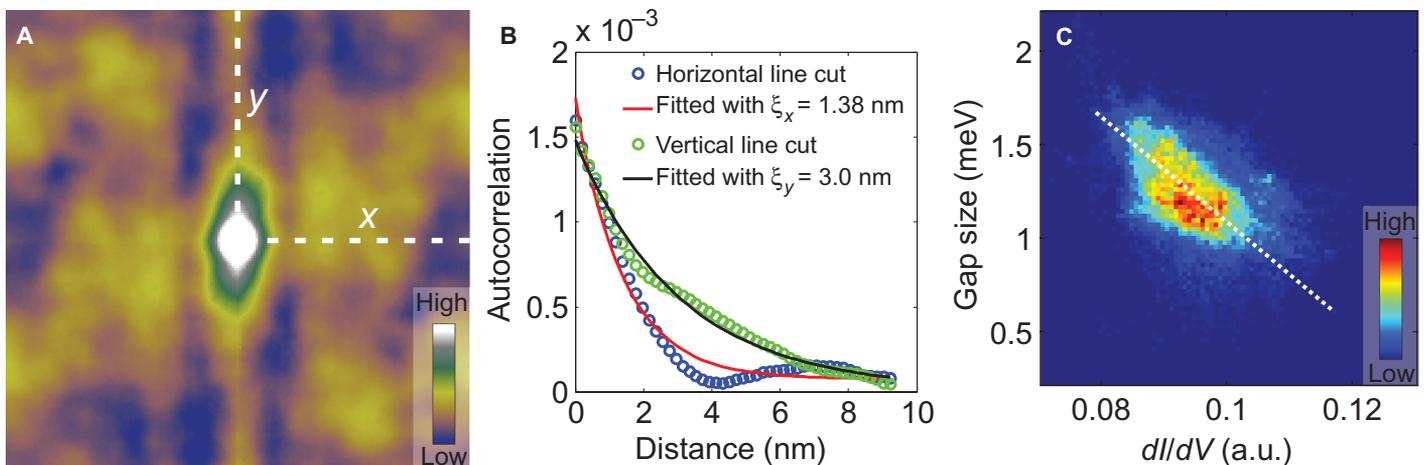

**Fig. 4. Relation between symmetry-breaking excitations and superconductivity.** (**A**) Autocorrelation of a gap map (38 × 38 nm²). (**B**) Line cuts (blue and green symbols) that are extracted along horizontal and vertical directions in (A) are fitted by an exponential decay function and yield decay lengths $\xi_x = 1.38$ nm and $\xi_y = 3.0$ nm. (**C**) Correlation plot between a low-pass filtered conductance map taken at $V = -4.5$ mV (see also fig. S5) and the local gap size in Fig. 1E. A clear anticorrelation (correlation coefficient ~−0.5) is found.

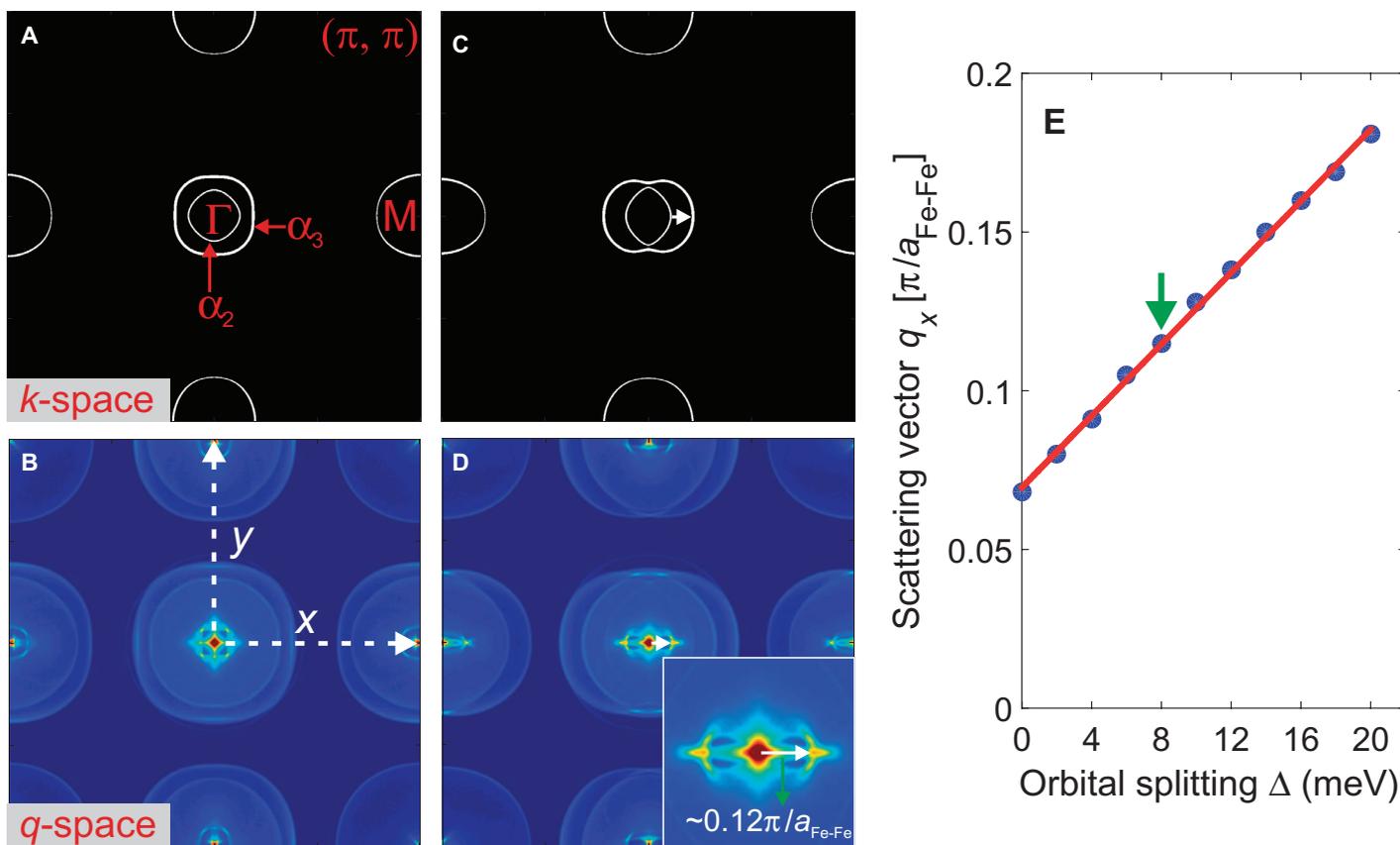

**Fig. 5. Comparison with the tight-binding model.** (**A**) Fermi surface of FeSe$_{0.4}$Te$_{0.6}$ as obtained from a tight-binding calculation in the normal state. (**B**) JDOS calculation corresponding to (A). (**C**) Fermi surface with orbital splitting (Δ) between $d_{xz}$ and $d_{yz}$ orbitals of 8 mV included. (**D**) Corresponding JDOS calculation. The dominant scattering vector near **q** = 0 is due to interband scattering between the α$_2$ and α$_3$ bands. The inset shows the magnified view of the central part. (**E**) Orbital splitting versus the dominant scattering vector **q**. The dominant scattering occurs on the hole pockets around the Γ point; a renormalization of 10 has been accounted for [which would be appropriate for the α$_3$ band (33)].

## DISCUSSION

In the iron chalcogenides, the parent compound FeTe does exhibit a magnetic and a structural phase transition but with a different ordering wave vector (27); hence, the symmetry breaking does not directly derive from the magnetostructural phase transition of the nonsuperconducting parent compound. Pure FeSe exhibits an orthorhombic phase transition and shows orbital modification above its structural transition (46) and anisotropic scattering patterns in STM (37, 38) in the orthorhombic phase similar to the ones we report here. Even upon suppression of the structural phase transition, which has been reported to occur around $x = 0.5$ in FeSe$_x$Te$_{1-x}$ (26), our results indicate that the electronic structure remains unstable against a reduction of symmetry from $C_4$ to $C_2$. This is in line with orbital excitations found in neutron scattering near $x = 0.5$ (47). In contrast to the iron pnictide compounds, where a similar anisotropy has been found in a number of observables, no spin density wave order has been reported for FeSe$_x$Te$_{1-x}$ with $x > 0.2$ (26, 44), supporting an interpretation in terms of local orbital ordering (40). Our data cannot discriminate between scenarios in which the electronic nematicity is induced by nematic fluctuations or driven by the structural distortion (48). Similar to our results, recent measurements show that nematicity not only develops at higher temperatures than where the orthorhombic distortion can be detected (14, 46) but also survives into the nonmagnetic superconducting regime of the phase diagram in BaFe$_2$(As$_{1-x}$P$_x$)$_2$ where the lattice structure is tetragonal (14). In this picture, the orbital excitations and nematic fluctuations detected in neutron scattering near $x = 0.5$ (47) become frozen out around defects and cause anisotropic scattering of charge carriers (17). A similar scenario could apply to anisotropic spin fluctuations, which have been detected in pnictides (49–51). An interpretation in terms of frozen fluctuations fits well with anisotropic states induced by cobalt defects in the Fe square lattice (16) and the resistivity anisotropy observed in cobalt-doped samples (15). Our samples have not been doped with impurities that would occupy the iron site; the major impurities are excess iron atoms that reside between the iron-chalcogenide planes. A possible explanation for the symmetry-breaking excitations seen in our sample is strain, for example, building up during cooldown because of thermal contraction, which might impose a small asymmetry in the orbital couplings. High sensitivity of iron-based superconductors to small strains has been previously reported (52). This interpretation would be consistent with our failure to observe domains of the symmetry-breaking states.

We observe an anticorrelation of the local size of the superconducting gap with a low-pass filtered map obtained at −4.5 mV. This anticorrelation can have a number of origins. Given that the dominant nonzero scattering vector at this energy is due to the symmetry-breaking states, this could indicate a competitive relationship with superconductivity; alternatively, because of the distortion of the Fermi surface, the super-conducting gap size might acquire a twofold anisotropy and hence regions with strong scattering patterns might exhibit an average gap size different from that in regions with low-intensity patterns. This interpretation would also rationalize the observed inhomogeneity of the superconducting gap size. Because of the small coherence length, the modulation in the local density of states due to the symmetry-breaking states can also lead to a change in the local gap size.

In summary, we have observed anisotropic QPI scattering along the Fe-Fe direction in Fourier transform maps, indicating $C_4$ symmetry breaking in a $FeSe_{0.4}Te_{0.6}$ superconductor. An analysis of the autocorrelation of gap maps shows evidence of an anisotropic superconducting coherence length along nearest-neighbor Fe-Fe directions, with a reduced coherence length in the direction of the dominant **q** vector of the symmetry-breaking excitations. The local intensity in a low-pass filtered map of the nematic excitations is found to anticorrelate with the local gap size. Comparison of the QPI data with tight-binding calculations including orbital splitting yields excellent agreement and allows us to estimate the splitting between $d_{xz}$ and $d_{yz}$ orbitals. Studies with in situ strain tuning (52, 53) may unravel the importance of strain for orbital ordering in iron chalcogenides.

## MATERIALS AND METHODS

### STM measurements

Measurements were performed with a homebuilt STM operating at a temperature below 2 K in cryogenic vacuum (54). The instrument allows for sample exchange and in situ sample cleavage at a temperature below 20 K. Maps were taken at several macroscopically different locations and with different slow scan directions. The STM tip was made from mechanically cut vanadium wire and cleaned by field emission on a single crystal gold sample in situ, ensuring that tunneling spectra were flat. The tip apex was also modified several times by contact to the surface during measurements.

Spectroscopic data were acquired as differential conductance through a lock-in amplifier with a small modulation voltage $V_{mod}$ added to the bias voltage. Maps consisting of up to more than 65,000 individual spectra were analyzed to extract QPI patterns.

### Sample growth and characterization

Single crystals of $FeSe_xTe_{1-x}$ were grown from self-flux. High-purity elements—99.99% Fe, 99.999% Se, and 99.999% Te—were used in the growth experiments. Se and Te were additionally purified by zone melting to reduce the amount of oxide impurities. The growth of single crystals was performed in evacuated double quartz ampoules at temperatures between 1000° and 1100°C. The final treatment of the samples was done at 410°C for 100 hours followed by quenching in ice water. The composition of the grown samples was investigated by EDX analysis. The phase content of the samples was analyzed by x-ray powder diffraction ($CuK_\alpha$ radiation, $\lambda = 1.540560$ Å) on crashed single crystals using a STADI-P powder diffractometer (STOE and CIE) with a position-sensitive detector [for detailed information, see the work of Tsurkan et al. (25)]. Additional characterization of the sample used for STM measurements is shown in fig. S1.

## SUPPLEMENTARY MATERIALS

Supplementary material for this article is available at http://advances.sciencemag.org/cgi/content/full/1/9/e1500206/DC1
Fig. S1. Magnetic susceptibility.
Note S1. Composition analysis from STM data.
Fig. S2. Composition analysis.
Note S2. QPI data processing.
Fig. S3. Unprocessed Fourier transforms and autocorrelation.
Fig. S4. Spectroscopic maps and QPI at 16 K.
Note S3. Correlation with symmetry-breaking states.
Fig. S5. Correlation with symmetry-breaking states.
Note S4. Tight-binding calculation.
Fig. S6. Details of the tight-binding model.
Note S5. JDOS calculation including orbital character.
Fig. S7. JDOS calculation including orbital character.
References (55, 56)

**Acknowledgments:** We are indebted to L. Boeri, F. Kruger, and P. Simon for discussion and suggestions and to V. Duppel for performing EDX analysis of the samples. **Funding:** S.C.W. and P.W. acknowledge partial support through the priority program SPP1458 (WA2510/1-1), and U.R.S. acknowledges support from the Alexander von Humboldt Foundation. P.W. further acknowledges support by the Engineering and Physical Sciences Research Council (grant EP/I031014/1). A.L., V.T., and J.D. acknowledge support by the German Science Foundation under grant no. DE1762/1-1 within SPP 1458. **Author contributions:** U.R.S and S.C.W. performed STM experiments. U.R.S., P.W., S.S., and S.C.W. analyzed the data. Samples were grown and characterized by J.D., V.T., and A.L. U.R.S. and P.W. wrote the manuscript. All authors discussed and contributed to the manuscript. P.W. designed the experiment. **Competing interests:** The authors declare that they have no competing interests. **Data and materials availability:** Underpinning data can be obtained from http://risweb.st-andrews.ac.uk/portal/.

Submitted 13 February 2015
Accepted 2 September 2015
Published 16 October 2015
10.1126/sciadv.1500206

**Citation:** U. R. Singh, S. C. White, S. Schmaus, V. Tsurkan, A. Loidl, J. Deisenhofer, P. Wahl, Evidence for orbital order and its relation to superconductivity in $FeSe_{0.4}Te_{0.6}$. Sci. Adv. **1**, e1500206 (2015).